# Spaser as Novel Versatile Biomedical Tool

Ekaterina I. Galanzha[1], Robert Weingold[1], Dmitry A. Nedosekin[1], Mustafa Sarimollaoglu[1], Alexander S. Kuchyanov[2], Roman G. Parkhomenko[3], Alexander I. Plekhanov[2], Mark I. Stockman[4], Vladimir P. Zharov[1]

[1] Winthrop P. Rockefeller Cancer Institute, Arkansas Nanomedicine Center, University of Arkansas for Medical Sciences, Little Rock, AR 72205, USA

[2] Institute of Automation and Electrometry of the Siberian Branch of the Russian Academy of Science, Koptyug Ave. 1, 630090 Novosibirsk, Russia

[3] Nikolaev Institute of Inorganic Chemistry of the Siberian Branch of the Russian Academy of Science, Lavrentiev Ave. 3, 630090 Novosibirsk, Russia

[4] Center for Nano-Optics and Department of Physics and Astronomy, Georgia State University, 29 Peachtree Center Ave., Atlanta, GA 30302, USA

**Fluorescence imaging and spectroscopy remain the most powerful tools for visualization with chemical and immunological specificity of labeled biomolecules, viruses, cellular organelles, and living cells in complex biological backgrounds[1]. However, a common drawback of fluorescence labels is that their brightness is limited by optical saturation and photobleaching[2]. As an alternative, plasmonic metal nanoparticles are very promising as optical labels with no photobleaching and low optical saturation at realistic exciting intensities[3] as was demonstrated in photoacoustic and photothermal sensing, imaging, and theranostics[4-9]. However, plasmonic nanoparticles have wide absorption spectra and are not fluorescent, which limits their spectral selectivity and multimodal functionality, respectively[6]. Here we demonstrate experimentally, both *in vitro* and *in vivo*, that spaser (surface plasmon amplification by stimulated emission of radiation)[10-12], also called nanolaser, provides unprecedented efficiency as a versatile tool in biomedical research and applications. This is due to the unique combination of intense near-monochromatic stimulated emission and strongly enhanced absorption, free of optical saturation. Using soluble and biocompatible uranine dye as a gain medium surrounding the gold nanocore as a plasmonic resonator, we demonstrate unprecedented spaser stimulated emission intensity ("giant spasing") and a narrow spectral width (0.8 nm), which are more than ~380-fold and**

**~30-fold, respectively, better than in quantum dots as the best conventional fluorescent nanoprobes. At the same time, the plasmonic spaser nanocore served as excellent photoacoustic and photothermal contrast agents for imaging and nanobubble–based theranostics of cancer cells. This makes the spasers, arguably, the best multifunctional, super-contrast, low-toxicity optical probes in biomedical research, especially with single-pulse excitation.**

Laser has revolutionized cell biology and disease diagnosis and treatment[13-15]. Laser-excited spontaneous fluorescence is a powerful diagnostic approach to study biomolecules and cells[1]. However, the fluorescence intensity (useful signal) saturates with excitation intensity while that of the background scattering does not[16]. Fundamentally different, stimulated emission does not saturate and can be used to boost sensitivity. This is the case in the so called biolaser[16] when biomolecules are placed directly in the laser cavity as the gain medium and pumped in order to amplify the signal in an optical cavity defined by mirrors. However, this approach requires an external optical cavity, and incorporates biological material into the gain medium, which limits the number of suitable biological objects and possible applications. Another promising strategy is the random laser, which uses a turbid gain medium with no mirrors[17-20]. However, applications of this technique to individual cells is problematic because it requires high intracellular concentration of the gain medium chromophores and strong multiple scattering inside the cell.

In contrast, the spaser[3] as a nanoscopic quantum generator of localized surface plasmons is a promising candidate for this mission. A spaser consists of a metal nanoparticle (NP) as the resonator surrounded by a nanoshell of the gain medium. Spasers were proposed in 2003 by Bergman and Stockman[10] and successfully demonstrated in 2009[21]. To date, many modifications of spasers have been developed[22-29]. The spasers can generate in a single mode with a very high spectral density and intensity. Potentially, the spaser have numerous applications in science, industry and, especially, in healthcare because their nanoscopic sizes are matching those of biomolecules. However, the spasers in biomedicine face challenges due to high metal optical losses, low solubility, and potential toxicity. Here we demonstrate, for the first time, a ~30-nm diameter spaser biocompatible with living cells and animals, which makes the brightest optical label with the highest signal-to-background-noise contrast for biomedicine. High local absorption of the spaser is an advantage in biomedical applications making the spaser one of the best and efficient multifunctional optical probes. We demonstrate that the developed techniques of photoacoustic (PA) and photothermal (PT) sensing and imaging with a capability for cancer theranostics[4-7] will tremendously benefit from the spasers.

We have synthesized a spaser consisting of 10±2 nm gold spherical NP surrounded by a 12±4 nm-thick silica shell doped with uranine (**Fig. 1a, Extended Data Fig. 1**), a low-toxicity disodium fluorescein soluble in water, which is widely used for medical diagnostics[30]. Two cancer cell lines, MDA-MB-231 (breast cancer) and MTLn3 (adenocarcinoma), were loaded by the spasers in concentration of $2\times10^{12}$ cm$^{-3}$ in solution through endocytosis. The presence of the spasers in the cells was verified through physical effects accompanying the interaction of pulsed pump (488 nm; 5-8 ns pulses) with the spasers: spontaneous fluorescence from the dye, stimulated emission (spasing), light scattering, and PT and PA phenomena (**Extended Data Fig. 2**). The spaser generation was initially confirmed in the solution for 2-mm, 120–μm (**Extended Data Fig. 3**), and 1-μm path lengths through the standard linear light-out vs. pump-in (L-P) dependence and spectral narrowing of the stimulated emission peak[12] with an increase of the pump intensity above a pronounced threshold of 5.7 MW/cm$^2$ (fluence ~40 mJ/cm$^2$ for 7 ns pulse width) (**Fig. 1b, inset**). The spasing was observed for a wide pump intensity range up to 400 MW/cm$^2$ accompanied by strong nonlinear PT-based effects related to transient vapor nano- and micro-bubble formation around overheated local absorbing regions[6, 31]. Above the spasing threshold, the L-P dependence in **Fig. 1b** is a straight line, as predicted[12], with emission spectral narrowing to 10-15 nm. At higher pump intensities above the threshold of bubble formation (~75 MW/cm$^2$), we observed further nonlinear enhancement of stimulated emission intensity and width narrowing (**Fig. 1b, d**). Irradiation of the spasers in solution in a thin (~1 μm) slide using a focused pump beam with a diameter of 1.1 μm and intensity of 300 MW/cm$^2$ allowed us to observe the highest ratio of the stimulated emission intensity to the spontaneous emission background of $1.3\times10^4$ ("giant spasing") and the narrowest emission peak of 0.8 nm, which were 18-fold and 7-fold, respectively, improvements over previous data[22-29]. Comparison of emission intensity of the spaser and quantum dots (QDs) – the best existing fluorescence probes[1, 5] – revealed 380-fold advantage of the spaser in brightness (**Fig. 1c, inset**). Taking into account the irradiated volume of the spasers (~$10^{-12}$ cm$^3$), our data suggest that just a single spaser can produce this effect. These data indicate that the spaser is an unprecedentedly superior optical probe for biomedicine.

At high pump intensities, we observed directional and even focused emission (**Fig, 1d, bottom; Extended Data Fig. 3c, d**) not described previously. We attribute this to a combination of scattering, refractive and thermal-lens effects (**SI Note 1**) due to transient vapor nano- and microbubbles around spasers (**Fig. 1d, middle, Extended Data Fig. 3b**).

Taking into account the strong optical absorption of the spasers, we have used them also as contrast agents in the PT and PA detection and imaging and in theranostics. We measured absorption and PA spectra of the spasers first at relatively low laser fluence. This shows the dominant dye absorption with the maximum absorption peak at 490 nm and also some contribution of the gold NP's absorption outside this peak (**Fig. 2a, Extended Data Fig. 4a**).

This suggests that, despite much higher absorption cross-section of the plasmonic gold NP [$\sigma_a$(NP)~$1\times10^{-12}$ cm$^{-2}$ for ~10-nm NPs at 525 nm] than of a dye molecule [$\sigma_a$(molecule)≈$1\times10^{-16}$ cm$^{-2}$ at 490 nm][1,9], the dye actually dominates absorption of the spaser near its absorption peak. The reason is that the number of dye molecules is large, $N_d$~$3\times10^3$, and the absorption cross section of each molecule is enhanced by local plasmonic fields by a factor ~$Q^2$, where $Q$~7 at 525 nm is the plasmonic quality factor for gold[12]. The total dye absorption cross section in the spaser is $\sigma_a$(dye)=$N_d Q^2 \sigma_a$(molecule) ≈$2.5\times10^{-10}$ cm$^2$, i.e., $\sigma_a$(dye)>>$\sigma_a$(NP).

In contrast, nonlinear PA spectroscopy[31] at high laser fluences revealed a maximum in the PA spectra matching the gold NP absorption (**Fig. 2a**). Comparison of PA signals from dye alone and the spasers revealed nonlinear signal increase from the spasers at high fluences associated with laser-induced nanobubble formation around overheated gold core (**Fig. 2b**).

We incubated spasers with viable cancer cells where they are actively internalized through endocytosis. The presence of the spasers in cells was verified by using dark field, PT, and fluorescent imaging (**Fig. 2c-e; Extended Data Fig. 6e, f**). Best image contrast, with 10-30-fold advantage compared to other imaging modalities, was demonstrated by the spasers (**Fig. 2e**). A relatively high PT image contrast (**Fig. 2c**) and strong PA effects (**Fig. 2a, b**) from the intracellular spasers suggest that the spasers can also be efficient multimodal contrast agents. In addition, laser-induced nanobubbles around strongly absorbing spaser's gold core can significantly amplify the PA signals (**Fig. 2b**), especially around clustered spasers (**SI Note 4**).

To explore the advantage of spasers' high brightness, we monitored emission from the spasers through a layer of human blood (**Fig. 3a**) or putting a blood layer on slide glass. First, for calibration purposes, we visualized cells on the blood surface **(Fig. 3b).** We were able to detect emission through a ~1 mm blood layer **(Fig. 3d)**, which would have been impossible with conventional fluorescence for depth >50-100 μm **(Fig.3c).** Thus, spaser provided a 10-fold improvement. Taking into account a signal-to-noise-ratio >30 in the obtained images **(Fig. 3d),** we anticipate a breakthrough in detection of individual cancer cells at an unprecedented depth ~1-3 cm.

Similar results were demonstrated *in vivo* by local injection of spasers in mouse ear tissue (**Fig. 3e**). The PA spectroscopy confirmed the presence of the spasers in the injected site through increased PA signal amplitudes (**Fig. 3f**) and typical for the spaser spectra **(Fig. 2a).** Pump irradiation demonstrated strong emission signal from the spasers through ear tissue (**Fig. 3g**) and quick (a few minutes) up taking of the spasers probably by lymphatics (**Extended Data Fig. 8**).

We estimated cell toxicity with the spasers at different concentrations and after laser irradiation. We observed no toxicity effects consistent with safety of uranine in medicine, life sciences, and water tracing[30]. Irradiation of cancer cells incubated with spasers with laser led to complete cells damage confirmed with cell viability assays, and visual observation of cell membrane damage and cell fragmentation **(Fig. 4a-c).** These data suggest potential of spasers as theranostic agents integrating noninvasive diagnosis at lower laser energy and cell killing at higher laser energy using just one or a few laser pulses **(Fig. 4c)**. The pump fluences used are close to laser safety standard for humans (20-100 mJ/cm$^2$ at 500-1100 nm)[32].

In conclusion, we have demonstrated the spasers as the highest-brightness, low-toxicity probes with narrow spectra. The spasers provide significant advantages in image contrast using just a single laser pulse, which is impossible with conventional probes (e.g., QDs). Moreover, we have showed multimodal capability of the spasers as PT and PA contrast agents for diagnosis and therapy with potential to use ultra-sharp PT-PA-plasmon resonances[31] (**Extended Data Fig. 5**) where just a single laser pulse is sufficient to kill cancer cells. The photobleaching of the spasers was negligible or insignificant at the pump intensities used (**Extended Data Fig. 9, Note 5**). The spasers emitted at least ~50-100 times brighter radiation than the dye alone due to the stimulated nature of the emission while the radiation of the spasers did not saturate with the pumping intensity. Note that possible spaser degradation at higher intensities is not important in the single pulse theranostic applications.

The spaser-based diagnostic and therapeutic platform established here can be further explored and developed. In particular, this may encompass extension of the spectral range to near–infrared tissue-transparency window (750-1100 nm), use narrow spaser emission for further increase of the detection sensitivity by narrow-band spectral filtering, potential spectral tunability for wavelength multiplexing without spectral overlap, and functionality enhancement by coating with appropriate protective polymer layers (e.g., PEG) and conjugation with antibodies and proteins for molecular targeting of cancer cells or infections[6].

**Supplementary Information** is at the end

**Acknowledgments** This work was supported in part by grants EB0005123 and CA131164 (both to VPZ). For MIS's work, the primary support was provided by a MURI Grant N00014-13-1-0649 from the US Office of Naval Research; an additional support was provided by a Grant No. DE-FG02-11ER46789 from the Materials Sciences and Engineering Division, Office of the Basic Energy Sciences, Office of Science, U.S. Department of Energy, and a Grant No. DE-FG02-01ER15213 from the Chemical Sciences, Biosciences and Geosciences Division, Office of the Basic Energy Sciences, Office of Science, U.S. Department of Energy.

**Author Contributions** EIA and VPZ designed the study; MIS proposed spaser type; ASK, RGP, and AIP sensitized and characterized spaser; RW, DAN, EIG and MS performed the measurements; VPZ and MIS wrote manuscript.

**Figures are below**

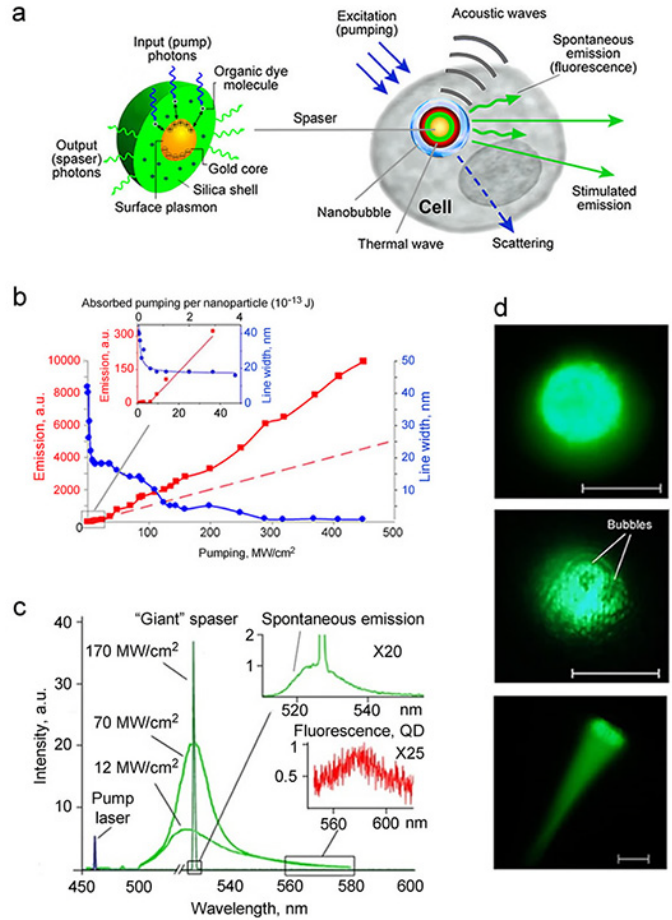

**Figure 1. Spaser for biological application. a,** Schematic of spaser as multifunctional intracellular nanoprobe. **b,** Stimulated emission in spaser suspension. Red: input-output (light out – pump [L-P]) curve (squares) of spasing. Inset shows curve at low laser intensity. Blue: emission linewidth of the spasing. The average SEM for each intensity is 21%. **c,** Radiation spectrum of spaser in suspension at 528 nm at different pump intensity (green curve), spontaneous emission (green, inset top), and quantum dot (QD) emission (x20) with maximum at 576 nm. Attenuated intensity of pump laser is shown by blue. **d,** Top**:** Spatially homogenous spaser's emission at a relatively low pump intensity (30 MW/cm$^2$, 120-μm thick spaser's suspension)**; Middle**: emission during the bubble formation around overheated spasers (150 MW/cm$^2$); **Bottom:** "directional" spaser emission in the presence of two large bubbles (200 MW/cm$^2$). Scale bars, 10 μm.

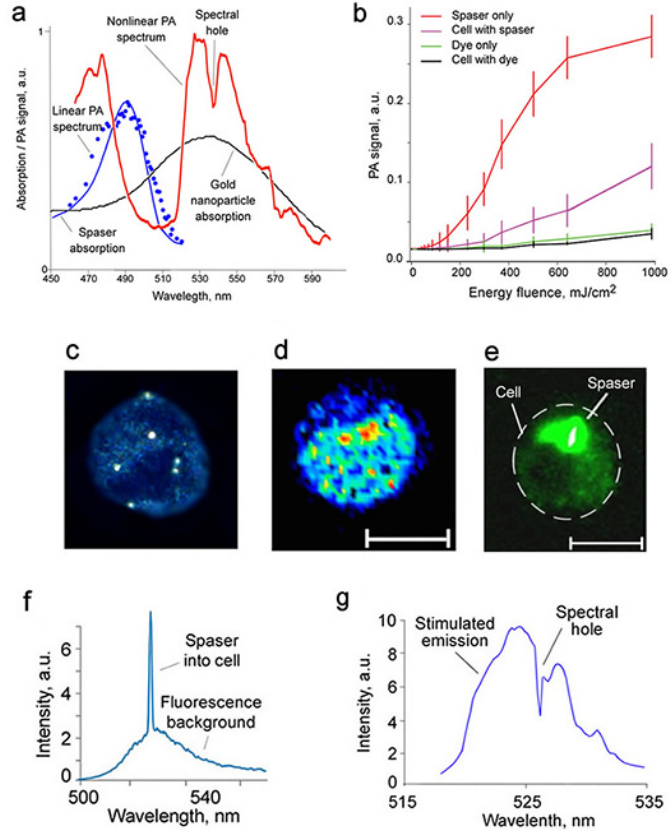

**Figure 2. Photoacoustic (PA) and photothermal (PT) spectral microscopy of spasers. a,** Absorption spectra, and linear and nonlinear PA spectra of spaser in suspension at laser energy fluence of 20 mJ/cm$^2$ and 150 mJ/cm$^2$, respectively. **b,** PA signal dependence on laser pump energy fluence for dye and spaser in suspension and into cells. **c-e,** Images of single cancer cell with spasers loaded through endocytosis (incubation time 10 min, temperature 37$^0$ C): (**c**)-scattering (dark field), (**d**) –photothermal (PT) ; (**e**)-conventional fluorescence; (**e**)- stimulated emission for local irradiated cell zone in background of conventional fluorescence image. **f,** Spectral peak from single cancer cell with spasers at relatively low energy fluence (80 mJ/cm$^2$). **g,** Spectral peak single cancer cell with spasers showing spectral hole burning in stimulated emission spectra at moderate energy fluence (135 mJ/cm$^2$). Scale bars, 10 μm.

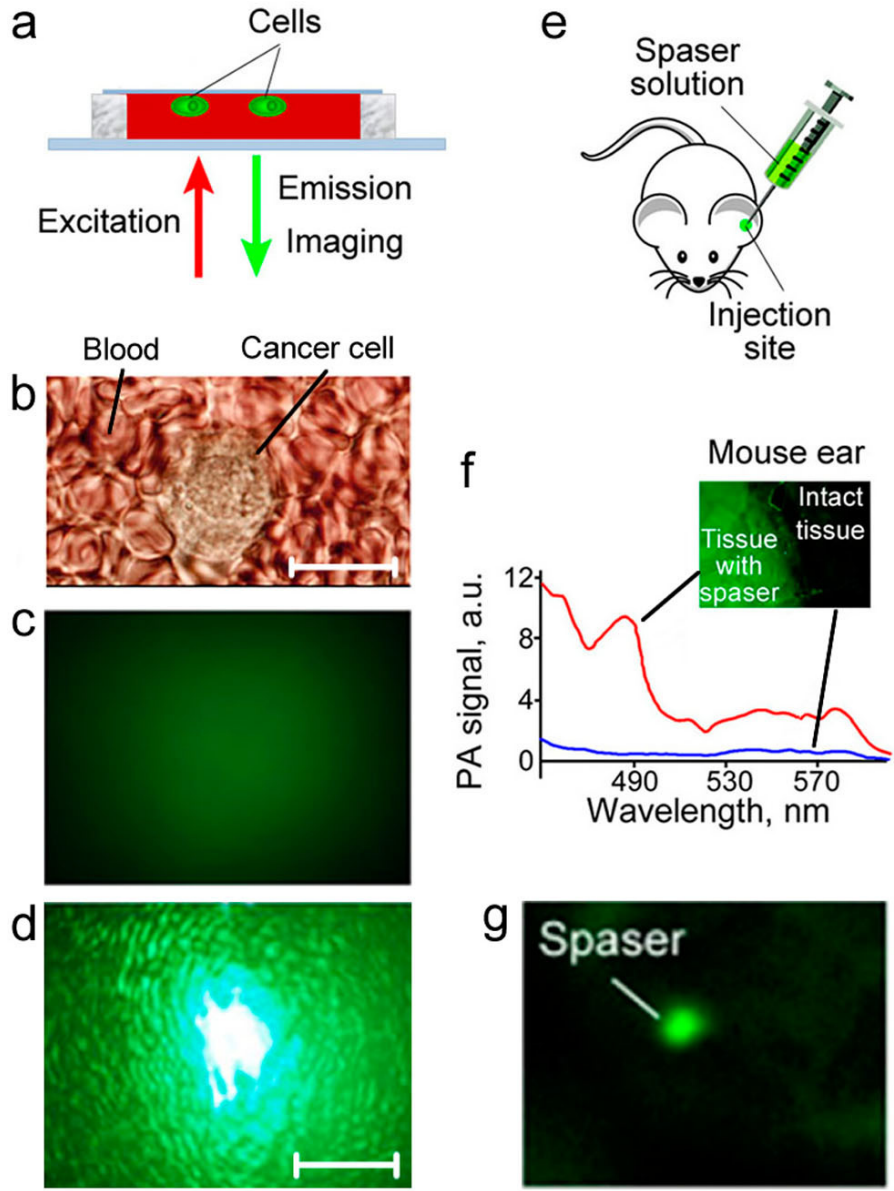

**Figure 3. Imaging of spasers in viable cells *in vitro* and in biotissue *in vivo*. a,** Schematic *in vitro*. **b,** Optical transmission image. **c, d,** Fluorescence imaging using conventional optical source (lamp) of blood with cancer cells at depth of ~1 mm **(top)** and spaser emission from single cancer cell at depth of 1 mm **(bottom). e,** Schematic of intradermal injection of spaser suspension into top layer of mouse ear tissue. **f,** PA identification of spasers in ear tissue using laser spectral scanning **(top).** Laser parameters: beam diameter 15 μm, fluence 20 mJ/cm$^2$. **g,** Spaser emission through ~250 μm ear tissue. Pump parameters: beam diameter: 10 μm**;** intensity, 30 MW/cm$^2$.

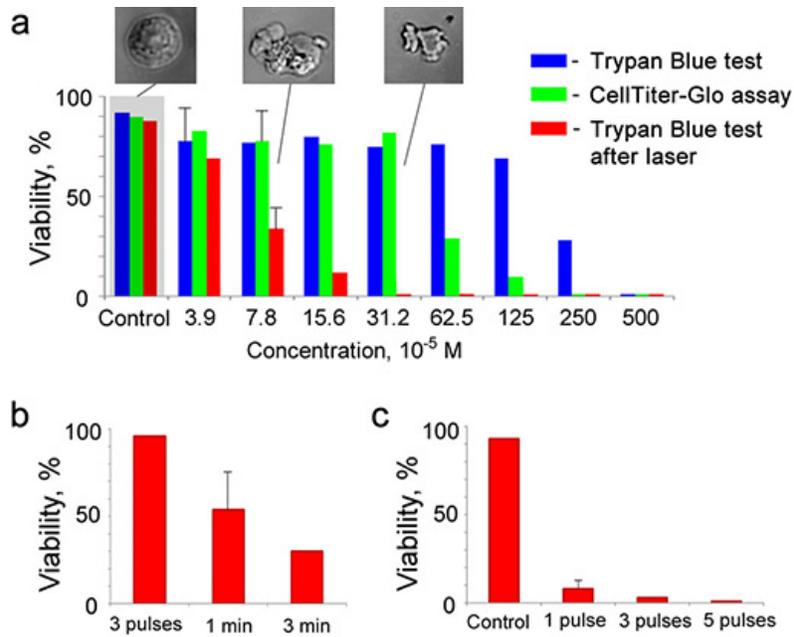

**Figure 4. Demonstration of spaser as theranostic agent**. **a,** Cell viability tests for different spaser concentration using two various kits without (blue, green) and after (red) laser irradiation (100 mJ/cm$^2$, 1 Hz, 3min). Inset: intact cell **(left)** and cells labeled with spasers at different concentration (middle and right) after laser irradiation. **b,** Viability test for concentration 15.6 x 10$^{-5}$ M as a function of laser exposure time (3s [3 pulses], 1 min, and 3 min); **c,** Viability test for concentration 15.6 x 10$^{-5}$ M as a function of laser pulse number (1, 3 and 5) showing that even single laser pulse at fluence of 500 mJ/cm$^2$ is sufficient for significant damage of cancer cells labeled by spasers. The average SD for each column is 15-20%.

## Methods

### Synthesis of spasers

Gold nanoparticles (NPs) were prepared by the standard method[33]. The next step was the modification of the gold surface by the addition of a freshly prepared aqueous solution of (3-aminopropyl) trimethoxysilane (5 µl, 1 mM) to the colloid under magnetic stirring. The gold particles were then individually coated in a sodium silicate shell by the addition of active silica (3 ml, 0.54 weight% sodium silicate solution) with pH10. After 1 day, the particles were transferred into ethanol (10 ml sol: 40 ml ethanol)[34]. The well-known Stober method was used to grow the shells[35]. The obtained colloid was mixed with the physiological solution of uranine $C_{20}H_{10}O_5Na_2$ ($2\times10^{-2}$ M) in equal volumes. Then the colloid was allowed to stand for 2 days to allow the infiltration of the dye into the shells. The concentration of the gold-shell NPs in a colloid was $\approx 2\times10^{12}$ cm$^{-3}$, calculated from the gold weight fraction measurements.

### Experimental setups

Spaser characterization was made with a multimodal photothermal (PT) and photoacoustic (PA) microscope described elsewhere[36]. Briefly, the PT microscopy module was built on technical platform of an Olympus IX81 inverted microscope (Olympus America, Inc.) using a tunable optical parametric oscillator (OPO; Opolette HR 355 LD, Opotek, Inc.) with the following parameters: tunable spectral range, 420–2,200 nm; pulse width, 5 ns; pulse repetition rate, 100 Hz; and fluence range, $0.1–10^4$ mJ/cm$^2$. Laser beams were focused into the sample using a 100× oil-immersion objective (DPlan 100, NA 1.25, Olympus, Inc.), 40× (Ach, NA 0.65, Olympus, Inc.), and a custom objective build using a single achromatic doublet lens (AC127-025-A-ML, Thorlabs, Inc.). In the PT thermal lens mode, pump (OPO) laser–induced temperature-dependent variations of the refractive index caused a collinear He-Ne laser probe beam (model 117A, Spectra-Physics, Inc.; wavelength, 633 nm; power, 1.4 mW) to defocus, and, hence, a reduction in the beam's intensity at its center. The probe laser light was collected after the sample using either a 100× water-immersion (LUMPlanFl 100, NA 1.00, Olympus, Inc.) or a 40× objective (Ph3 DL, NA 0.55, Nikon Inc) and was detected by a photodiode (PDA36A, Thorlabs, Inc) with a 50-µm pinhole (referred to as PT signals). The PT signals demonstrate an initial peak associated with rapid, picosecond-scale heating of dyes or NPs and with a slower, nano- to microsecond-scale, exponential tail corresponding to target cooling. In the nonlinear mode, laser-induced nanobubbles around overheating strongly-absorbing zones (e.g., NPs) lead to the appearance of sharp negative peaks associated with refraction and scattering of the probe beam on the nanobubbles (**Extended Data Fig. 2,** upper right). In the confocal scheme, the plane of the pinhole is fixed one Rayleigh distance from the probe-beam waist. The PT images are constructed by acquiring the PT signals from a sample as it undergoes scanning in *x-y* dimensions using a two-dimensional stage (H117 ProScan II, Prior Scientific, Inc.) with the scanning step of 0.25-1 µm. For 3-D imaging, successive PT images are acquired in parallel *x-y* planes distributed along the *z* axis. Focusing along the z axis was performed by moving the microscope objective axially (a minimum of 50-nm step). The PT signals were recorded using a 200-MHz analog-to-digital converter board (National Instruments Corp., PCI-5152, 12-bit card, 128 MB of memory) and analyzed by custom software (LabVIEW 8.5 complex; National Instruments). A Dell Precision 690 workstation provided signal acquisition/processing, synchronization of the excitation laser, and translation-stage control. Resolution is determined by the microscope objective itself (e.g., ~0.7 µm at 20×, NA 0.4; and ~250 nm at 100×, NA 1.25). In the PA module, the PA signals from an ultrasound transducer (unfocused XMS-310 with a 10-MHz frequency band, or focused V316-SM with a 20-MHz frequency band; both from Panametrics) and an amplifier (model 5662 or 5678, Panametrics) were recorded using customized software. Fluorescence imaging, added to verify PT and PA data, was performed with a cooled color CCD camera (DP72, Olympus). Navigation of the laser beams for the *in vitro* study was controlled with a high-resolution (~300 nm) transmission digital microscopy module (TDM). In the enhanced dark-field microscopy mode, an enhanced illuminator (CytoViva Inc., Auburn, AL) replaced custom light collection condenser used for the PT imaging. The illuminator consisted of a CytoViva 150 condenser and a fiber optic light guide connected to a Solarc 24W metal halide light source (Welch Allyn, Skaneateles Falls, NY). A 100× oil objective with an iris (Olympus UPlanAPO fluorite, N.A. 1.35–0.55) and a color camera (DP72, Olympus America Inc.) were used to acquire sample images. This scheme significantly increases the scattering contrast of the NPs inside cells.

**Spectroscopic measurement of spaser emission**

To study the stimulated emission, samples were loaded in a cuvette (with oblique walls to eliminate feedback) of 2 mm path length and pumped at wavelength λ = 488 nm with 7-ns pulses from an optical parametric oscillator (Solar LP601) lightly focused into a 1-mm spot. With increasing pump. the emission peak appeared at 528-530 nm (**Fig. 1d, Extended Data Fig. 1b**, curve 2), which became narrower once the pumping energy exceeded a critical threshold value (**Extended Data Fig. 1b**, curve 3). For spectroscopic measurements, we used a fiber-optic-based spectrometer OceanOptics USB4000 with optical resolution of $\Delta\lambda \sim 0.1$ nm (FWHM) or AvaSpec-2048 TEC-FT-2 [$\Delta\lambda \sim 0.7$ nm (FWHM)].

In **Figure 1c**, inset shows the intensity of this peak as a function of pumping energy, yielding an input–output curve with a pronounced threshold characteristic of lasers. The threshold intensity is equal to 5.7 MW/cm$^2$. The ratio of the intensity of this laser peak to the spontaneous emission background increased with increasing pumping energy. At the same time, there was a narrowing of the lasing spectrum at the threshold (**Extended Data Fig. 1b**, curve 3).

**Cells**

Human breast cancer cells (MDA-MB-231, American Type Culture Collection [ATCC]) were cultured according to the vendor's specifications. MTLn3 adenocarcinoma cells, originally isolated by Neri and Nicolson (Institute for Molecular Medicine, Huntington Beach, CA) were maintained in α-minimum essential medium (Invitrogen/Life Technologies) with 5% fetal bovine serum and penicillin–streptomycin (Invitrogen/Life Technologies). Viable cells were resuspended in phosphate buffered saline (PBS) at concentration of $\sim 1 \times 10^6$ cells per 1 mL and labeled with spaser (10 min, 37$^0$ C).

**Animal and human study**

According to University of Arkansas for Medical Sciences Institutional Animal Care and Use Committee approved protocols, the experiment involved a nude mouse model purchased from Harlan Sprague-Dawley (Indianapolis, IN). The animals were anesthetized by isoflurane and placed on a microscopic stage using ultrasound gel for acoustic matching of the ultrasound transducer and the ear tissue. The positions of ear structures, pump beam, and injected site were visualized by optical and fluorescence imaging.

Fresh blood was obtained from healthy donors in heparinized tubes in accordance with protocols approved by the UAMS Institutional Review Board.

**Study of cell viability**

The impact of the laser irradiation on live cells in the presence of the spasers was estimated with standard cell viability assays: trypan blue and CelTiter-Glo kits according to manufacture procedures. The highest spaser concentration in the suspension (5 x $10^{-3}$ M) (Fig. 4) was sequentially dilated two times.

**Statistical analysis**

Results are expressed as the means plus/minus the standard deviation and the confidence interval of at least three independent experiments (P ¼ 0.95). Statistica 5.11 (StatSoft, Inc.), MATLAB 7.0.1 (MathWorks), and LabVIEW (National Instruments) were used for the statistical calculations. Data were summarized as the mean, standard deviation (SD), relative standard deviation, median, and full range. The error bars in figures represent SD. In the selected figures, we only indicated the average SEM/SD in the caption to provide more clear data presentation.

**Extended Data Figures are below**

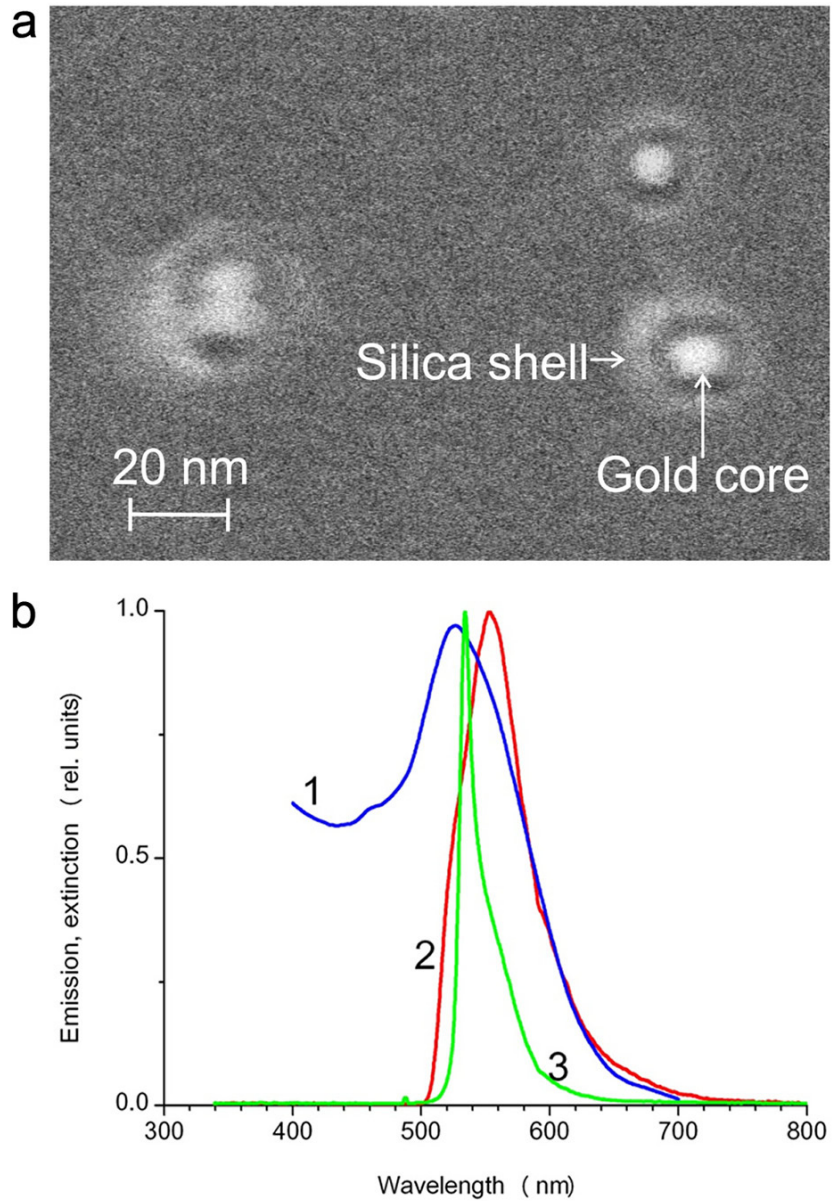

**Extended Data Figure 1. a,** TEM images of single and clustered spasers. **b**, Spectroscopic results. (1) Normalized: extinction of gold nanoparticles (blue curve) as core of spaser (average diameter <D> = 10 nm and concentration c = 2·$10^{12}$ cm$^{-3}$). (2) Luminescence of the dye (red curve) at the low pump intensity (2 MW/cm2). (3) Stimulated emission (green curve) of the gold nanoparticles in dye-doped shell above the threshold (5.7 MW/cm2).

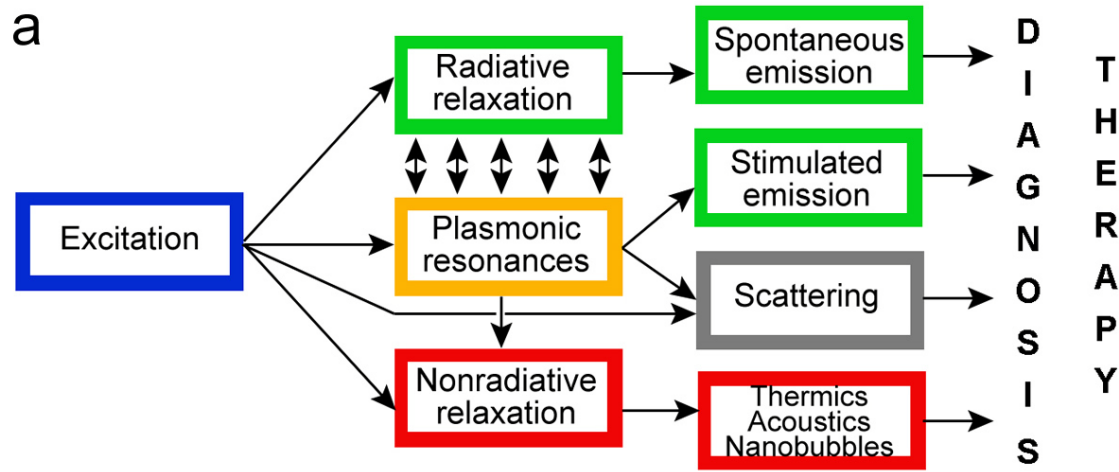
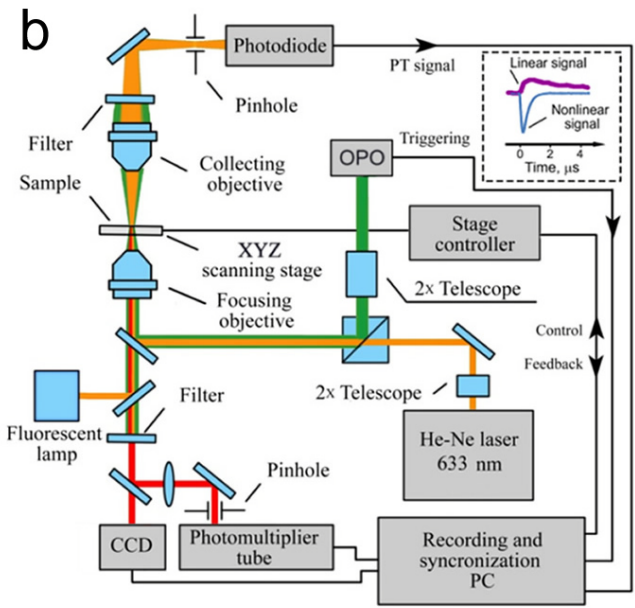
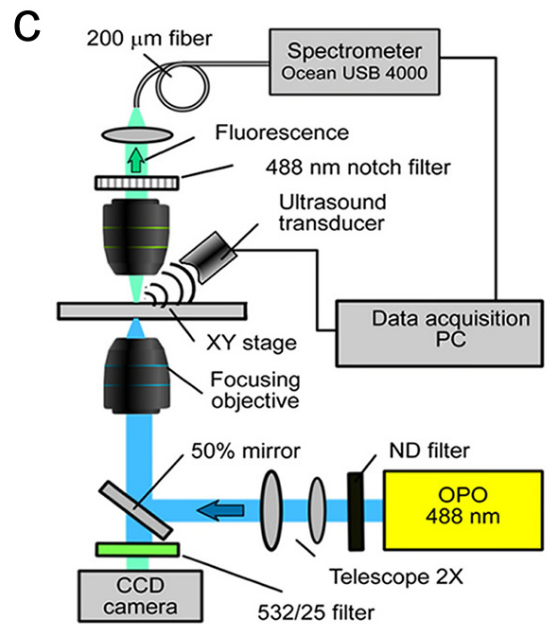

**Extended Data Figure 2. Schematics of the physical effects in spaser and experiments setups**. **a,** Key physical phenomena in cell with spaser under laser-spaser interaction. **b,** Integrated photothermal (PT), photoacoustic (PA), and fluorescence (with conventional lamp) microscopy. **c,** Integrated PA and fluorescence microscopy using Ocean USB 4000 spectrometer.

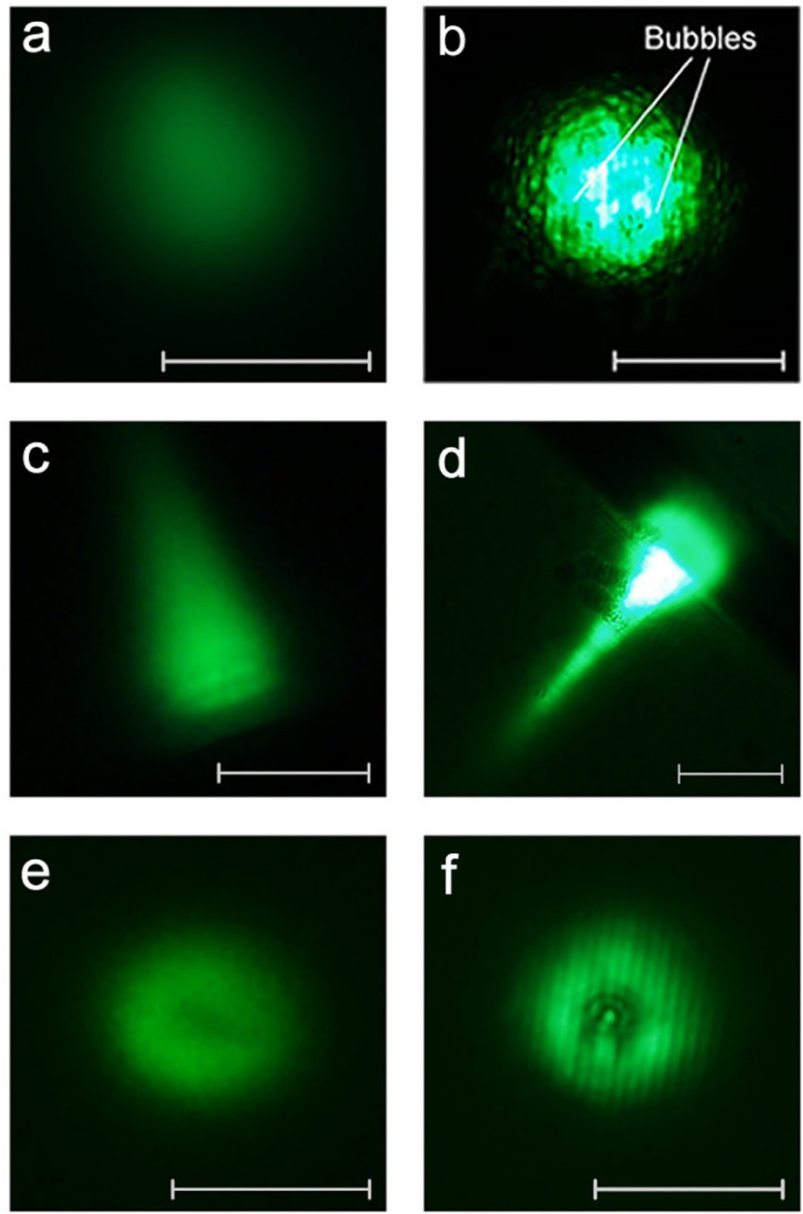

**Extended Data Figure 3. Emission of spaser during irradiation (pumping) of spaser suspension in 120-μm thick slide with 10-μm pump beam at different pump intensity. a,b**, Fluorescence images at pump intensity below (5 MW/cm$^2$) and above (60 MW/cm$^2$) threshold (5.7 MW/cm$^2$), respectively. **c,** Directional emission during bubble formation around clustered spasers (pump intensity, 250 MW/cm$^2$). **d,** Focusing of spaser emission during bubble formation (pump intensity, 270 MW/cm$^2$). **e,** Spatial hole burned in the center of pump beam (300 MW/cm$^2$). **f,** Heterogeneous spatial hole burning (350 MW/cm$^2$). Scale bar, 10 μm.

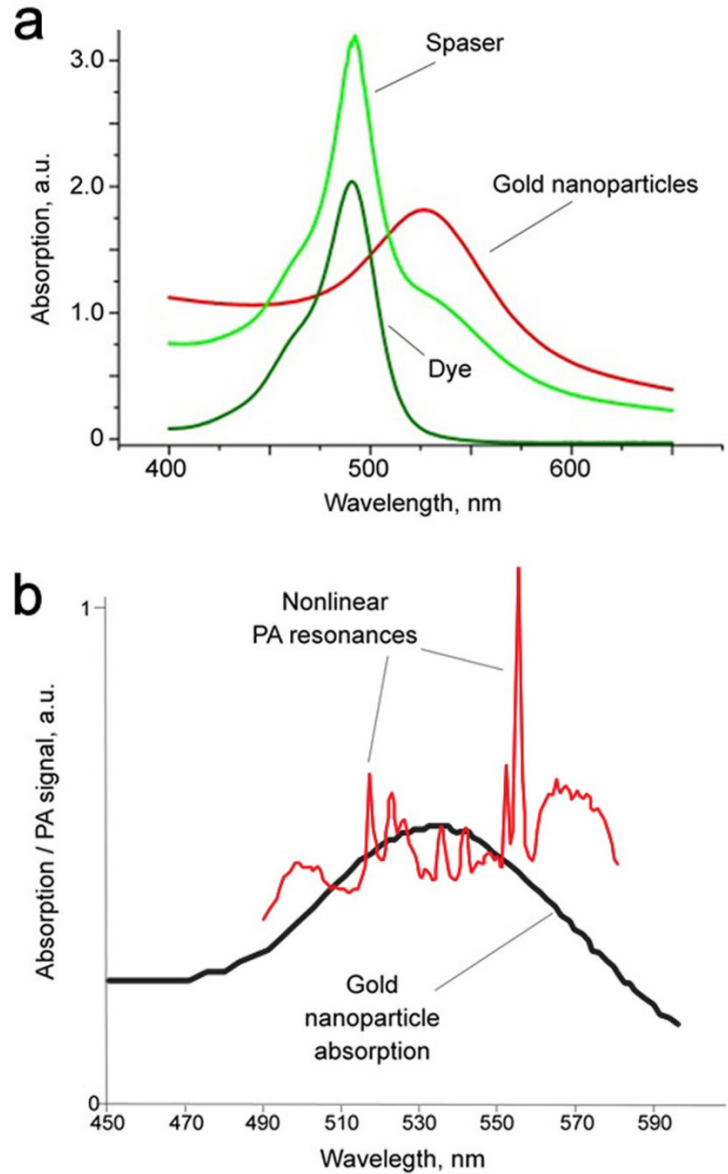

**Extended Data Figure 4. a**, Absorption spectra of gold nanoparticles (red curve), dye alone (dark green), and spaser (light green). **b**, Absorption spectrum (black) and nonlinear PA spectrum (red) of gold nanoparticles constituting gold core of spaser. PA data show the splitting of plasmon resonance at 530 nm in two -blue and red shifted PA peaks allowing better spaser identification in a complex biological environment.

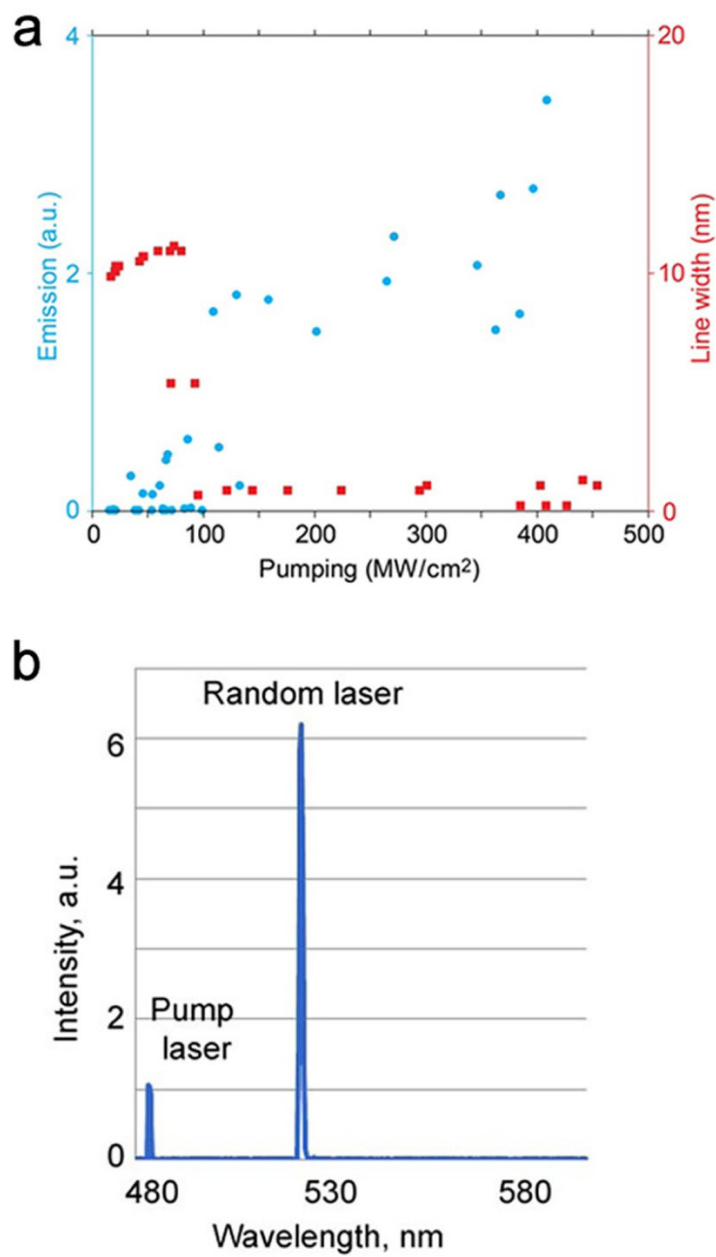

**Extended Data Figure 5. Stimulated emission in dye solution (random laser). a**, Blue: input-output (light out – pump [L-P]) curve (circles) of spasing in dye solution only. This lasing was unstable with a threshold associated with pump-induced nano- and microbubble formation around the dye aggregates, Red: emission line width of the lasing (squares). **b,** Example of generation of spectrally narrow (~2 nm) stimulated emission peak in the dye solution alone (Random laser effect) at the pump intensity of 300 MW/cm$^2$ at 488 nm.

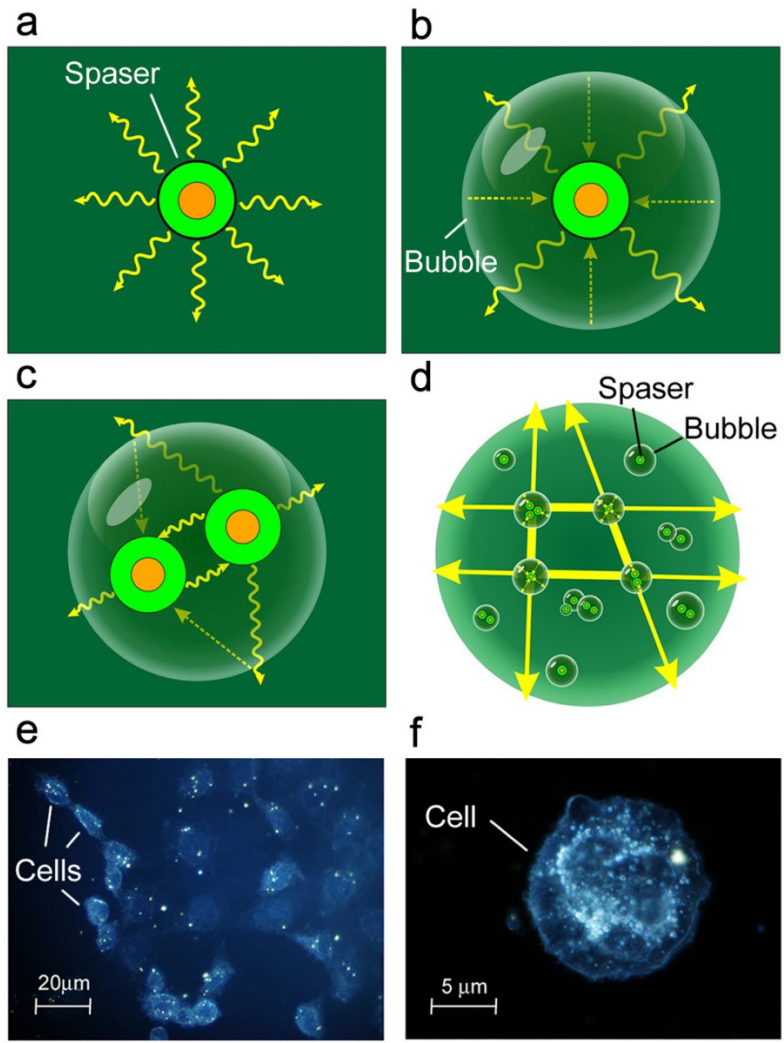

**Extended Data Figure 6. Concept of "random nano-bubble laser". a,** Schematic of single spaser with no bubble around; **b,** Single spaser with bubble with scattering (feedback) from internal nanobubble surface; **c,** Two spasers, when emission from one spaser goes to another and is reflected from external nanobubble surface; **d,** Multiple bubbles when laser-induced small transient vapor nanobubbles occurring around overheated absorbing zones at high energy fluence can scatter out light emitted in the process of photoluminescence in such a way that feedback is introduced to the system and coherent and incoherent random lasing can be observed. **e-f,** Dark-field microscopic images of spasers (as scattering centers) inside multiple (left) and single (right) cancer cells.

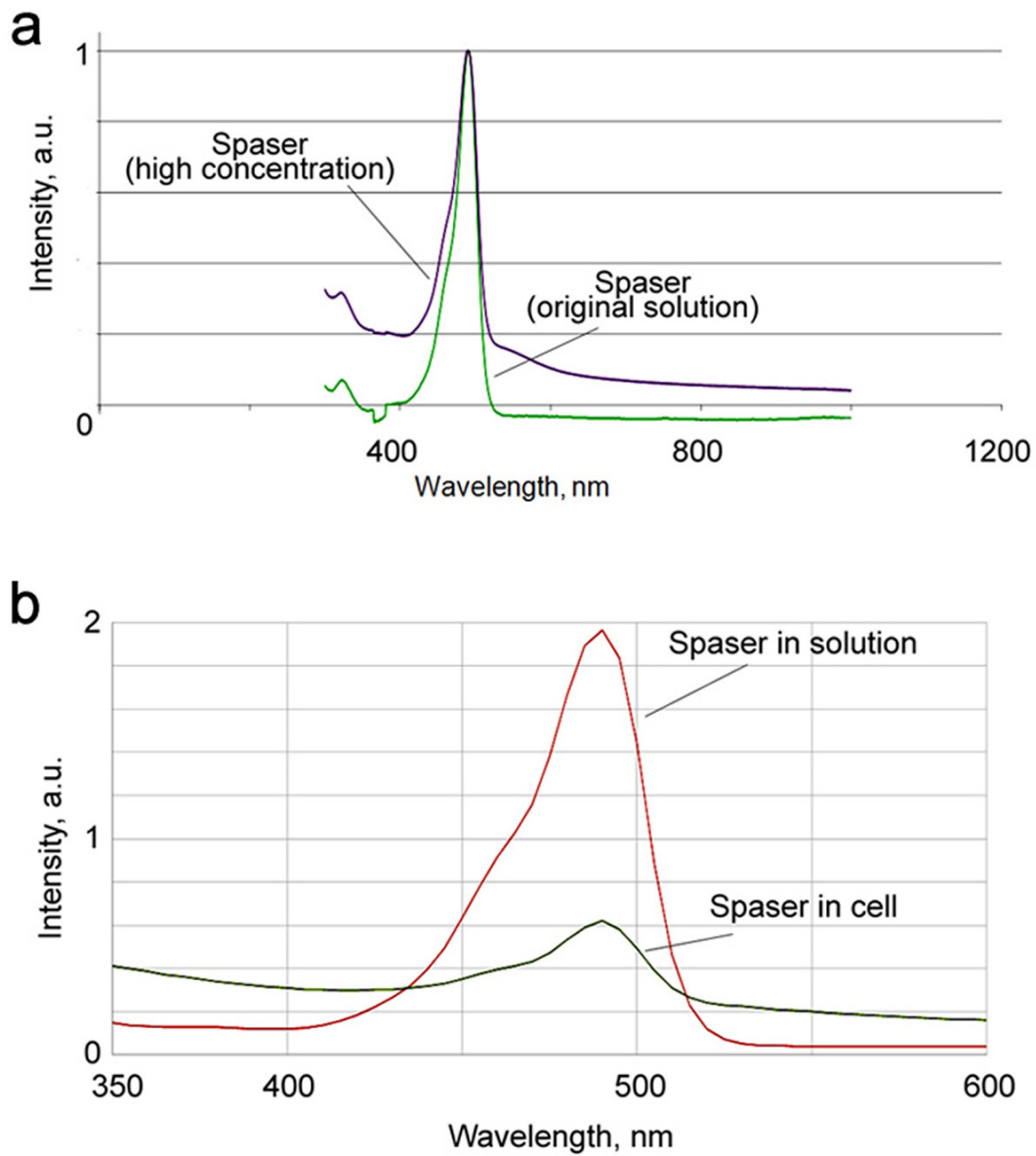

**Extended Data Figure 7. a**, Absorption spectra of the spaser in suspension (green) and after concentrating by centrifugation (violet). **b**, Absorption spectra of the spaser in suspension (red) and in cancer cells (green).

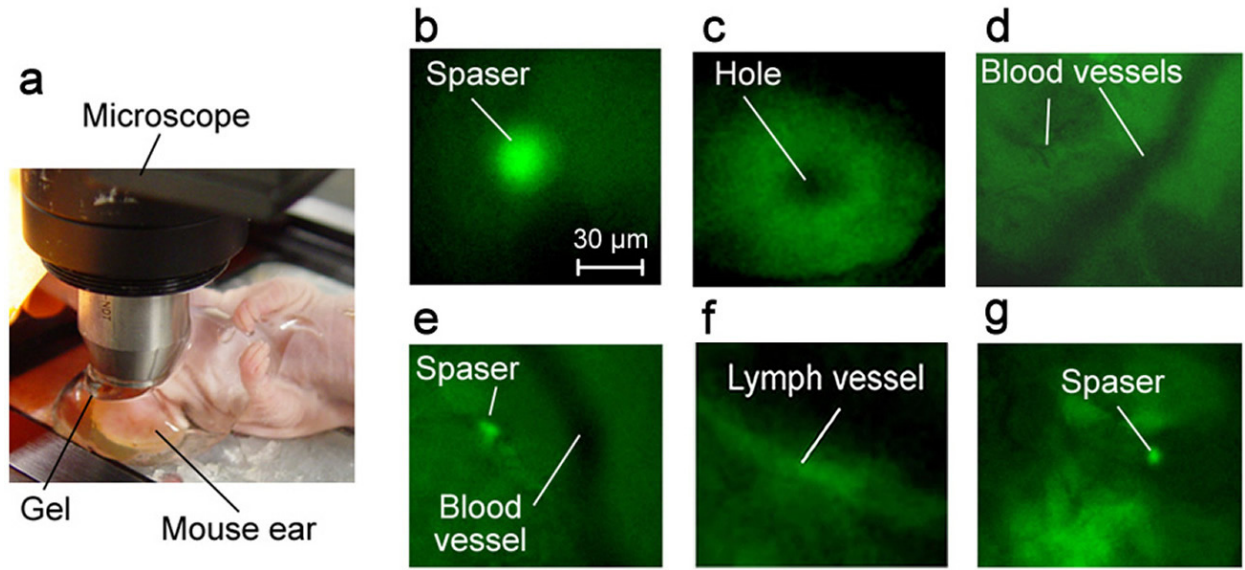

**Extended Data Figure 8. Imaging of spasers in biotissue *in vivo* (mouse model). a,** Schematic of PA identification of spasers in mouse ear. **b,** Spaser emission through mouse ear at a relatively low intensity of focused pump beam (30 MW/cm$^2$). **c,** Spatial hole burning at high pump intensity of pump beam (70 MW/cm$^2$) with a beam diameter of 30 µm. **d**, Imaging of ear vasculature. **e-g,** Various patterns of spaser emission through the ear tissue.

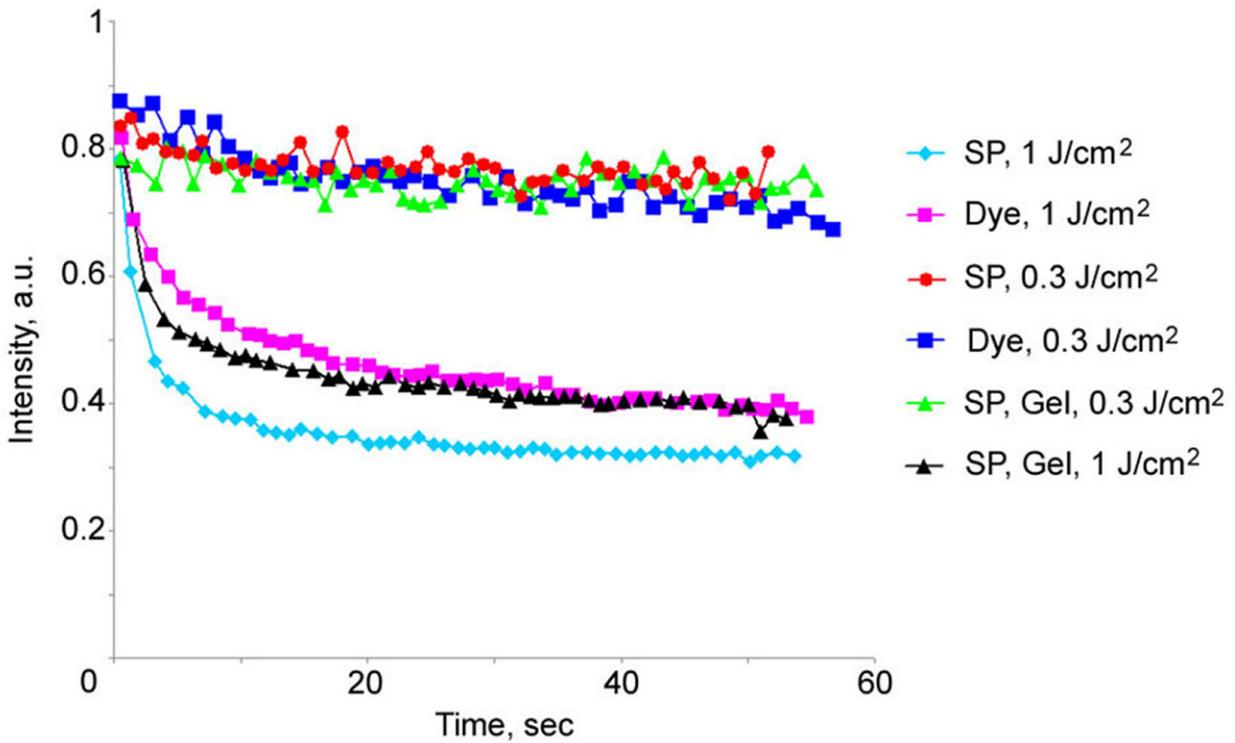

**Extended Data Figure 9.** Fluorescence intensity from spasers and dye alone at high (3 J/cm$^2$) and low (0.3 J/cm$^2$) laser energy fluence as function of time in suspension and in gel. The average SEM for each time point is 16%.

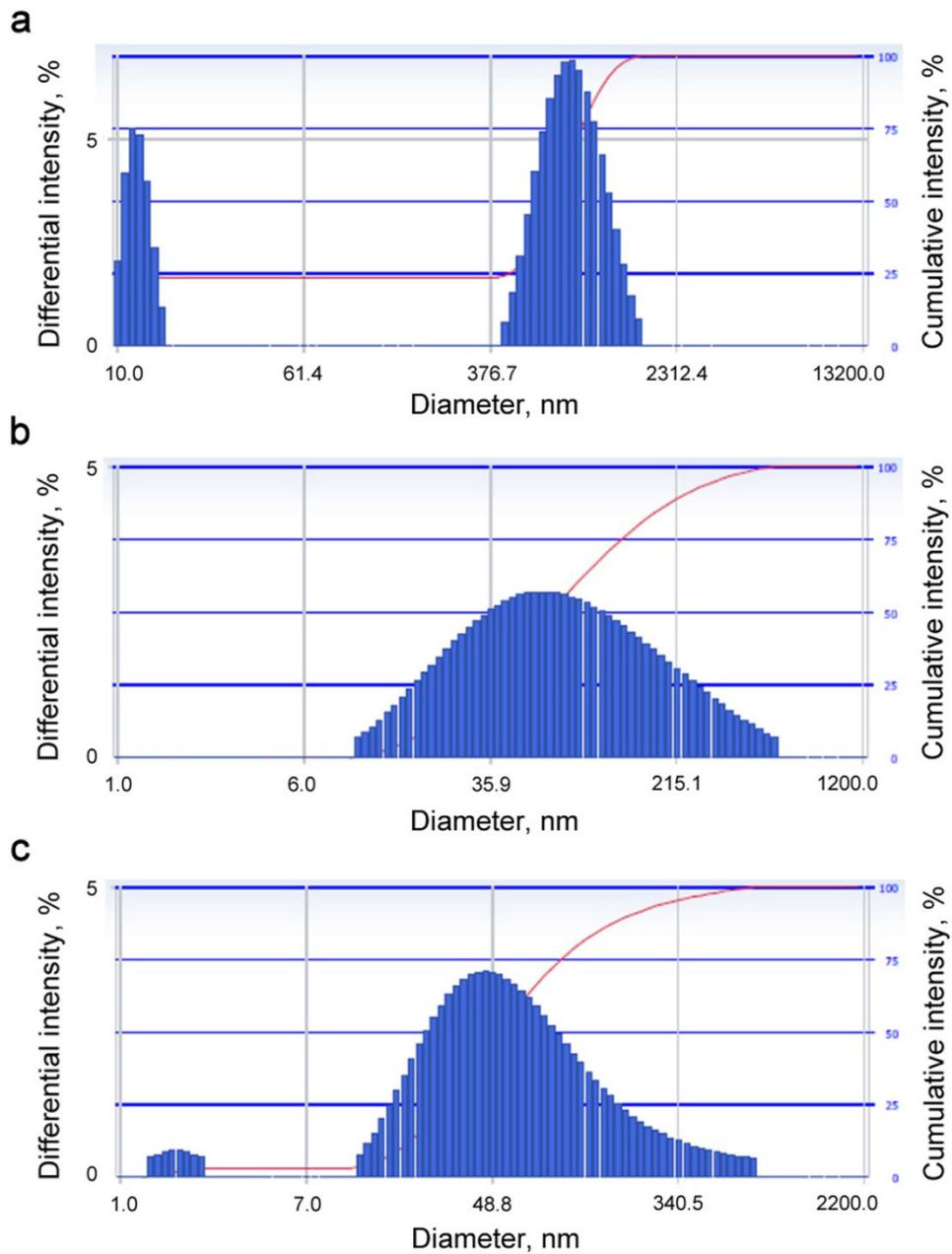

**Extended Data Figure 10.** Differential intensity distribution and cumulative intensity of the total scattered intensity due to: **a**, NPs (inclusions) in the filtered physiological solution of uranine after one week. **b**, Gold NPs in a silica shell in the aqueous solution; **c**, Gold NPs in a silica shell in physiological solution of uranine.

**Supplementary Notes**

**Note 1**

At relatively low pump intensity, (6-50 MW/cm$^2$), change of the spatial orientation of the slide with a spaser suspension with respect to the pump beam direction within an angle of ± 30$^0$ revealed almost isotropic emission distribution within the accuracy of 10-15%. This finding suggests an advantage of the spasers as optical probes because the signal is independent of the spatial orientation for the spasers inside cells with respect to both the direction of the pump laser and the axis of emission collection. At a higher pump intensity, we observed interesting optical phenomena such as directional emission (**Fig.1d, bottom; Extended Data Fig. 3c**), emission focusing (**Extended Data Fig. 3c**) and spatial hole burning (**Extended Data Fig. 3e,f**). These phenomena can be associated with overheating of strongly absorbing local zones (e.g., spaser clusters) leading to formation of transient vapor nano- and microbubble around these zones (**Fig. 1d, middle; Extended Data Fig. 3b**). The size and life time of these bubbles measured with various optical methods[37,38] were in the broad range of 20 nm to 10 μm and 1ns to 5 μs, respectively, depending on pump intensity. Strong refractive, scattering, and thermal lens effects in highly localized heated areas and especially bubbles can be responsible for the light concentrating and redirecting. High local transient temperature around the gold NPs (up to several hundred degrees centigrade) led to spaser photobleaching of photothermal (PT) origin accompanied by appearance of dark spot ("spatial hole burning") (**Extended Data Fig. 3e**) or heterogeneous light distribution (**Extended Data Fig. 3f**) in the center of Gaussian pump beam where intensity is maximum (300-350 MW/cm$^2$). We believe that these effects can be used to study nonlinear optical effects in the spaser and optimization of its biological applications using low pump intensity for diagnostics and an increased intensity for theranostics.

**Note 2**

Spectral specificity of plasmonic labels is limited to only two simultaneous colors because of relatively broad absorption bands of NPs (e.g., 50-100 nm for gold nanospheres). The discovery of ultrasharp PT and PA resonances in plasmonic NPs with a spectral width down to 1 nm (i.e., ~100-fold narrower than linear absorption spectra) could overcome this problem[31]. The corresponding physical mechanism is based on nanobubble-associated nonlinear signal enhancement. Specifically, a shift of the laser wavelength toward the absorption center leads to increased energy absorption that raises the temperature above a well-defined nanobubble-formation threshold, accompanied by a significant (10–50-fold) signal enhancement. As a result, spectrally dependent signal amplification leads to sharpening of PT and PA resonances only near the centers of absorption peaks. At a laser fluence much above the nanobubble threshold, a laser will burn (e.g., through NP melting, thus decreasing absorption) a spectral hole with two red- and blue-shifted sharp PT and PA resonances formed[39,40] (**Extended Data Fig. 4b**). The discovery of the peak-splitting phenomenon accompanied by dramatic narrowing of the split red- and blue-shifted peaks have been recently termed *Zharov splitting*[40]. It was also indicated that the spectral peak splitting with increased laser energy is analogous to *supercritical bifurcation* in nonlinear dynamic systems, and that both phenomena raise the surprising possibility of its being universal behavior in nonequilibrium systems with many potential biological applications[40].

**Note 3**

At a high pump intensity, we observed narrow (1-2 nm) emission peaks at 524-525 nm in dye solution alone at the highest concentration used. This emission intensity and spectra demonstrated dependence on laser fluence similar to that of the spaser, (**Extended Data Fig. 5**) but at higher (70-120 MW/cm$^2$) threshold. The optical components with water instead of the dye did not produce these effects. High resolution fluorescence microscopy revealed some heterogeneous image structures associated with dye aggregates. Measurement of the PA signals from dye solution revealed nonlinear signal increase at high fluence associated with laser-induced nanobubble formation around overheated irradiated zones (**Fig. 2b**). The filtration of dye solution led to complete disappearance of the emission peaks above the background. Based on these data and "sharp" threshold typical for phase transition. we hypothesize that the observed phenomena can be associated with stimulated emission of the random laser due to scattering light on dye aggregates and especially on small transient vapor laser-induced nanobubbles in which reflection increase 10-50 times compared to liquid state **(Extended Data Fig. 6).** To our knowledge, such high level of emission of the random laser amplified by nano-bubbles ("random nano-bubble laser") and its super-narrow spectral emission peak was observed for the first time. However, we did not observe the random lasing in a cell due to both low local intracellular dye concentration compared to the bulk dye solution and too small cell size for a random resonator. Nevertheless, the spatial and temporal characteristics of a random nanobuble-enhanced laser formed in large-scale biosamples (e.g., in animal skin) can provide information about structural properties of the tissue and, therefore, may be applied for diagnosis of diseases, such as dermatological disorders or skin cancer.

**Note 4**

Our obtained data may suggest the presence of spaser clusters inside cells[41], which can enhance stimulated emission through plasmonic coupling. A comparison of the absorption spectra of the dye and spasers shows dominant dye absorption with the maximum absorption peak at 490 nm and some contribution of the gold NP's absorption outside of this peak (on the wings) in the range $\leq$475 nm and $\geq$520 nm (**Fig. 2a, Extended Data Fig. 4a)** especially after concentrating the spasers **(Extended Data Fig. 7a)**. A comparison of the absorption spectra of the dye and the spasers in suspension and inside cells reveals slight red-shifted absorption and its enhancement outside of the dye peak (**Extended Data Fig. 7b**) as an indicator of gold- NP plasmonic coupling, although silica shells prevent more profound plasmonic coupling between the gold cores. The measuring of the PA signals from dye and spasers alone and from cells loaded with the spasers or the dye as a function of the pump laser fluence at 532 nm demonstrated more profound nonlinear signal enhancement from the spasers compared to the dye solution (**Fig. 2b**). Moreover, despite lower concentration of the spasers inside the cells compared to the initial concentration in suspension before incubation with the cells, the signal amplitude levels are similar, which suggests stronger nanobubble formation around the clustered spasers inside the cells than in suspension around individual spasers **(Fig. 2b).**

Note 5

Measurement of fluorescent intensity from the spasers and dye alone in solution and in gel at different pump laser fluence indicates no photobleaching at a relatively high fluence of 0.1-0.5 $J/cm^2$. The spasers were more resistant to photobleaching than the dye alone suggesting that silica matrix stabilizes the dye and the faster transfer of energy from the dye to the spaser core protects it from the degradation. In gel with more restricted spatial motion photobleaching was more profound (**Extended Data Fig. 9**). Spaser-produced emission intensity inside cells **(Fig. 2e)** was lower compared to emission in suspension **(Fig. 1c)** and spectrally wider (3-5 nm) sometimes containing a few (2-4) closely located peaks with total width of 5-15 nm due to lower spaser concentration inside the cells than in suspension before incubation with the cells. At low pump energy, the cells remain alive even after prolonged laser exposure up to 1,000 pulses at 100 nJ/pulse. At higher pump energy, we occasionally observed spectral hole (negative peaks) in the stimulated emission spectra (**Fig. 2g**) that could be associated with quenching phenomena especially in spaser clusters.

**Note 6**

Using a 0.9% NaCl aqueous solution may lead to aggregation of dye molecules, especially, at high concentration. The dye aggregates enhanced by the laser-induced bubble formation[41] served as scattering centers that made the random lasing possible (see Note 2). These effects were eliminated by solution filtering. Nevertheless, the process of aggregation of dye molecules as well as the spasers in a dye solution can occur with time even in filtered solutions. Using photon correlation spectroscopy (DelsaNano C, Beckman Coulter, Inc.), we determined the hydrodynamic size of the inclusions of about 590 nm (Polydispersity Index (P.I.) = 0.382) in the filtered solution after one week (**Extended Data Fig. 10**). The pure aqueous solution of gold NPs in a silica shell gives hydrodynamic size of 58 nm (P.I. = 0.313) (i.e., the physical size is about 1.5 times less) (**Extended Data Fig. 10a**), but after mixing this solution with the physiological solution of uranine, the average size became 190 nm after one week (P.I. = 0.096) (**Extended Data Fig. 10b**). It led also to formation of the aggregates (clusters) consisting of two or three spasers. This effect is apparently associated with the ability NaCl of the electrolyte to cause NP aggregation. Although spaser clusters can provide both amplification of stimulated emission and theranostic efficiency by reducing the bubble formation threshold, in most current studies we used solution filtering immediately before application. The stability of the colloidal system can be also improved by spaser coating, for example, by polyoxyethyleneglycol.

**References**

37. Lin, C. P. & Kelly, M. W. Cavitation and acoustic emission around laser-heated microparticles. *Appl. Phys. Lett*. **72**, 2800-2807 (1998).